\documentstyle[11pt,epsfig]{article}
\begin{document}
\title{\Large\bf  Towards the assignment for the $4\,^1S_0$ meson nonet }

\author{\small De-Min Li\footnote{E-mail: lidm@zzu.edu.cn}~ and Shan Zhou\\
\small   Department of Physics, Zhengzhou University, Zhengzhou,
Henan 450052, People's Republic of China}
\date{\today}
\maketitle
\vspace{0.5cm}

\begin{abstract}

The strong decays of the $\pi(2070)$, $\eta(2010)$, $\eta(2100)$,
$\eta(2190)$, and $\eta(2225)$ as the $4\,^1S_0$ quark-antiquark
states are investigated in the framework of the $^3P_0$ meson
decay model. It is found that the $\pi(2070)$, $\eta(2100)$, and
$\eta(2225)$ appear to be the convincing $4\,^1S_0$ $q\bar{q}$
states while the assignment of the $\eta(2010)$ and $\eta(2190)$
as the $4\,^1S_0$ isoscalar states is not favored by their widths.
In the presence of the $\pi(2070)$, $\eta(2100)$, and $\eta(2225)$
being the members of the $4\,^1S_0$ meson nonet, the $4\,^1S_0$
kaon is phenomenologically determined to has a mass of about 2153
MeV. The width of this unobserved kaon is expected to be about 197
MeV in the $^3P_0$ decay model.

\end{abstract}

\vspace{0.5cm}
 {\bf Key words:} mesons, $^3P_0$ model

 {\bf PACS numbers:}14.40. Cs, 12.39.Jh

\newpage

\baselineskip 24pt

\section{Introduction}
\indent \vspace*{-1cm}

 From PDG2006\cite{pdg2006}, the $1\,^1S_0$ meson nonet ($\pi$, $\eta$,
$\eta^\prime$, and $K$) as well as the $2\,^1S_0$ members
[$\pi(1300)$, $\eta(1295)$, and $\eta(1475)$] has been well
established. In Ref.\cite{eta1835}, we suggested that the
$\pi(1800)$, $K(1830)$, together with the $X(1835)$ and
$\eta(1760)$ observed by BES Collaboration\cite{x1835,BES17602}
constitute the $3\,^1S_0$ meson nonet. More recently, we argued
that the $\eta(2225)$ with a mass of $(2240^{+30+30}_{-20-20})$
MeV and a width of $(190\pm 30^{+40}_{-60})$ MeV observed by the
BES Collaboraton\cite{BES08} could be the $4\,^1S_0$ $s\bar{s}$ in
Ref. \cite{eta2225} where the other members of the $4\,^1S_0$
meson nonet were not discussed. In the present work, we shall
address the possible assignment for the $4\,^1S_0$ meson nonet.

With the assignment of the $\eta(2225)$ as the $s\bar{s}$ member
of the $4\,^1S_0$ meson nonet, one can expect that both the
isovector and another isoscalar members of the $4\,^1S_0$ meson
nonet should be lighter than the $\eta(2225)$. Experimentally, in
the mass region 2000-2225 MeV, the pseudoscalar states $\pi(2070)$
[Mass: $2070\pm 35$ MeV, Width: $310^{+100}_{-50}$ MeV],
$\eta(2010)$ [Mass: $2010^{+35}_{-60}$ MeV, Width: $270\pm 60$
MeV], $\eta(2100)$ [Mass: $2103\pm 50$ MeV, Width: $187\pm 75$
MeV], and $\eta(2190)$ [Mass: $2190\pm 50$ MeV, Width: $850\pm
100$ MeV] are reported\cite{pdg2006}. Theoretically, some
predicted values for the $\pi(4\,^1S_0)$ mass are 2.15 GeV
 by QCD sum rules\cite{qcdsump1,qcdsump2}, 2.009 GeV
by the spectrum integral equation\cite{spectrum} , 2.193 GeV by a
covariant quark model\cite{covqm}, 2.039 GeV by a relativistic
independent quark model\cite{riqm}, and 2.07 GeV by Regge
phenomenology\cite{reggeph}. In addition, the mass of the third
radial excitation of the $\eta$ is predicted to be about 2.267 GeV
by a covariant quark model\cite{covqm} or 2.1 GeV by Regge
phenomenology\cite{reggeph}. The $\pi(2070)$ mass is similar to
the predicted $\pi(4\,^1S_0)$ mass, and all the masses of the
$\eta(2010)$, $\eta(2100)$, and $\eta(2190)$ are close to the
predicted mass range of the third radial excitation of the $\eta$.
Only the mass information of these states is insufficient to
classify them.  The main purpose of this work is to discuss
whether these reported pseudoscalar states can be assigned as the
members of the $4\,^1S_0$ meson nonet or not by investigating
their decay properties in the $^3P_0$ meson decay model.

The organization of this paper is as follows. In section 2, the
brief review of the $^3P_0$ decay model is given (for the detailed
review see {\sl e.g.}
Refs.\cite{3p0rev1,3p0rev2,3p0rev3,3p0rev4}.) In sections 3 and 4,
the decay widths of the $\pi(2070)$, $\eta(2010)$, $\eta(2100)$,
$\eta(2190)$, and $\eta(2225)$ as the $4\,^1S_0$ $q\bar{q}$ state
are presented. The decay widths of the $4\,^1S_0$ kaon are
predicted in section 5, and the summary and conclusion are given
in section 6.

\section{ The $^3P_0$ meson decay model}
\indent \vspace*{-1cm}

 The $^3P_0$ decay model, also known as the quark-pair creation
model, was originally introduced by Micu\cite{micu} and further
developed by Le Yaouanc et al.\cite{3p0rev1}. The $^3P_0$ decay
model which ( in several variants) is the standard model for
strong decays at least for mesons in the initial state, has been
widely used to evaluate the strong decays of
hadrons\cite{3p00,3p0y,3p0x,3p0x1,3p0x2,3p01,3p02,3p03,quarkmass,3p04},
since it gives a good description of many of the observed decay
amplitudes and partial widths of the hadrons. The main assumption
of the $^3P_0$ decay model is that strong decays take place via
the creation of a $^3P_0$ quark-antiquark pair from the vacuum.
The new produced quark-antiquark pair, together with the
$q\bar{q}$ within the initial meson regroups into two outgoing
mesons in all possible quark rearrangement ways, which corresponds
to the two decay diagrams as shown in Fig.1 for the meson decay
process $A\rightarrow B+C$.

\begin{figure}[hbt]
\begin{center}
\epsfig{file=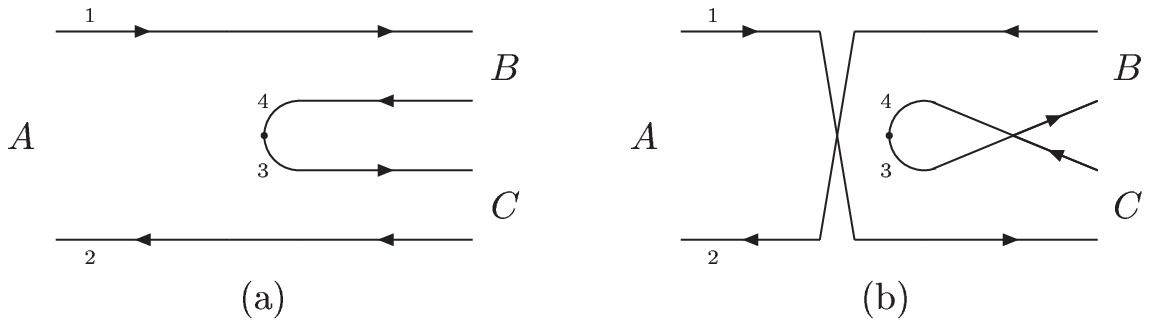,width=8.0cm, clip=}
\vspace*{0.5cm}\vspace*{-1cm}
 \caption{\small The two possible diagrams contributing to $A\rightarrow B+C$ in the $^3P_0$
 model.}
\end{center}
\end{figure}

The transition operator $T$ of the decay $A\rightarrow BC$ in the
$^3P_0$ model is given by
\begin{eqnarray}
T=-3\gamma\sum_m\langle 1m1-m|00\rangle\int
d^3\vec{p}_3d^3\vec{p}_4\delta^3(\vec{p}_3+\vec{p}_4){\cal{Y}}^m_1(\frac{\vec{p}_3-\vec{p}_4}{2})\chi^{34}_{1-m}\phi^{34}_0\omega^{34}_0b^\dagger_3(\vec{p}_3)d^\dagger_4(\vec{p}_4),
\end{eqnarray}
where $\gamma$ is a dimensionless parameter representing the
probability of the quark-antiquark pair $q_3\bar{q}_4$ with
$J^{PC}=0^{++}$ creation from the vacuum, $\vec{p}_3$ and
$\vec{p}_4$ are the momenta of the created quark $q_3$ and
antiquark  $\bar{q}_4$, respectively. $\phi^{34}_{0}$,
$\omega^{34}_0$, and $\chi_{{1,-m}}^{34}$ are the flavor, color,
and spin wave functions of the  $q_3\bar{q}_4$, respectively. The
solid harmonic polynomial
${\cal{Y}}^m_1(\vec{p})\equiv|p|^1Y^m_1(\theta_p,\phi_p)$ reflects
the momentum-space distribution of the $q_3\bar{q}_4$ .

For the meson wave function, we adopt the mock meson
$|A(n^{2S_A+1}_AL_{A}\,\mbox{}_{J_A M_{J_A}})(\vec{P}_A)\rangle$
defined by\cite{mock}
\begin{eqnarray}
|A(n^{2S_A+1}_AL_{A}\,\mbox{}_{J_A M_{J_A}})(\vec{P}_A)\rangle
&\equiv& \sqrt{2E_A}\sum_{M_{L_A},M_{S_A}}\langle L_A M_{L_A} S_A
M_{S_A}|J_A
M_{J_A}\rangle\nonumber\\
&&\times  \int d^3\vec{p}_A\psi_{n_AL_AM_{L_A}}(\vec{p}_A)\chi^{12}_{S_AM_{S_A}}\phi^{12}_A\omega^{12}_A\nonumber\\
&&\times  |q_1({\scriptstyle
\frac{m_1}{m_1+m_2}}\vec{P}_A+\vec{p}_A)\bar{q}_2
({\scriptstyle\frac{m_2}{m_1+m_2}}\vec{P}_A-\vec{p}_A)\rangle,
\end{eqnarray}
where $m_1$ and $m_2$ are the masses of the quark $q_1$ with a
momentum of $\vec{p}_1$ and the antiquark $\bar{q}_2$ with a
momentum of $\vec{p}_2$, respectively. $n_A$ is the radial quantum
number of the meson $A$ composed of $q_1\bar{q}_2$.
$\vec{S}_A=\vec{s}_{q_1}+\vec{s}_{q_2}$,
$\vec{J}_A=\vec{L}_A+\vec{S}_A$, $\vec{s}_{q_1}$ ($\vec{s}_{q_2}$)
is the spin of $q_1$ ($q_2$), $\vec{L}_A$ is the relative orbital
angular momentum between $q_1$ and $q_2$.
$\vec{P}_A=\vec{p}_1+\vec{p}_2$,
$\vec{p}_A=\frac{m_1\vec{p}_1-m_1\vec{p}_2}{m_1+m_2}$. $\langle
L_A M_{L_A} S_A M_{S_A}|J_A M_{J_A}\rangle$ is a Clebsch-Gordan
coefficient, and $E_A$ is the total energy of the meson $A$.
$\chi^{12}_{S_AM_{S_A}}$, $\phi^{12}_A$, $\omega^{12}_A$, and
$\psi_{n_AL_AM_{L_A}}(\vec{p}_A)$ are the spin, flavor, color, and
space wave functions of the meson $A$, respectively. The mock
meson satisfies the normalization condition
\begin{eqnarray}
\langle A(n^{2S_A+1}_AL_{A}\,\mbox{}_{J_A M_{J_A}})(\vec{P}_A)
|A(n^{2S_A+1}_AL_{A}\,\mbox{}_{J_A
M_{J_A}})(\vec{P}^\prime_A)\rangle=2E_A\delta^3(\vec{P}_A-\vec{P}^\prime_A).
\end{eqnarray}
The $S$-matrix of the process $A\rightarrow BC$ is defined by
\begin{eqnarray}
\langle BC|S|A\rangle=I-2\pi i\delta(E_A-E_B-E_C)\langle
BC|T|A\rangle,
\end{eqnarray}
with
\begin{eqnarray}
\langle
BC|T|A\rangle=\delta^3(\vec{P}_A-\vec{P}_B-\vec{P}_C){\cal{M}}^{M_{J_A}M_{J_B}M_{J_C}},
\end{eqnarray}
where ${\cal{M}}^{M_{J_A}M_{J_B}M_{J_C}}$ is the helicity
amplitude of $A\rightarrow BC$. In the center of mass frame of
meson $A$, ${\cal{M}}^{M_{J_A}M_{J_B}M_{J_C}}$ can be written as
\begin{eqnarray}
{\cal{M}}^{M_{J_A}M_{J_B}M_{J_C}}(\vec{P})&=&\gamma\sqrt{8E_AE_BE_C}
\sum_{\renewcommand{\arraystretch}{.5}\begin{array}[t]{l}
\scriptstyle M_{L_A},M_{S_A},\\\scriptstyle M_{L_B},M_{S_B},\\
\scriptstyle M_{L_C},M_{S_C},m
\end{array}}\renewcommand{\arraystretch}{1}\!\!
\langle L_AM_{L_A}S_AM_{S_A}|J_AM_{J_A}\rangle\nonumber\\
&&\times\langle L_BM_{L_B}S_BM_{S_B}|J_BM_{J_B}\rangle\langle
L_CM_{L_C}S_CM_{S_C}|J_CM_{J_C}\rangle\nonumber\\
&&\times\langle 1m1-m|00\rangle\langle
\chi^{14}_{S_BM_{S_B}}\chi^{32}_{S_CM_{S_C}}|\chi^{12}_{S_AM_{S_A}}\chi^{34}_{1-m}\rangle\nonumber\\
&&\times[f_1I(\vec{P},m_1,m_2,m_3)+(-1)^{1+S_A+S_B+S_C}f_2I(-\vec{P},m_2,m_1,m_3)],
\end{eqnarray}
with $f_1=\langle
\phi^{14}_B\phi^{32}_C|\phi^{12}_A\phi^{34}_0\rangle$ and $f_2=
\langle\phi^{32}_B\phi^{14}_C|\phi^{12}_A\phi^{34}_0\rangle$,
corresponding to the contributions from Figs. 1 (a) and 1 (b),
respectively, and
\begin{eqnarray} I(\vec{P},m_1,m_2,m_3)&=&\int
d^3\vec{p}\,\mbox{}\psi^\ast_{n_BL_BM_{L_B}}
({\scriptstyle\frac{m_3}{m_1+m_2}}\vec{P}_B+\vec{p})\psi^\ast_{n_CL_CM_{L_C}}
({\scriptstyle\frac{m_3}{m_2+m_3}}\vec{P}_B+\vec{p})\nonumber\\
&&\times\psi_{n_AL_AM_{L_A}}
(\vec{P}_B+\vec{p}){\cal{Y}}^m_1(\vec{p}),
\end{eqnarray}
where $\vec{P}=\vec{P}_B=-\vec{P}_C$, $\vec{p}=\vec{p}_3$, $m_3$
is the mass of the created quark $q_3$.

The spin overlap in terms of Winger's $9j$ symbol can be given by
\begin{eqnarray}
&&\langle
\chi^{14}_{S_BM_{S_B}}\chi^{32}_{S_CM_{S_C}}|\chi^{12}_{S_AM_{S_A}}\chi^{34}_{1-m}\rangle=\nonumber\\
&&\sum_{S,M_S}\langle S_BM_{S_B}S_CM_{S_C}|SM_S\rangle\langle
S_AM_{S_A}1-m|SM_S\rangle\nonumber\\
&&\times(-1)^{S_C+1}\sqrt{3(2S_A+1)(2S_B+1)(2S_C+1)}\left\{\begin{array}{ccc}
\frac{1}{2}&\frac{1}{2}&S_A\\
\frac{1}{2}&\frac{1}{2}&1\\
S_B&S_C&S
\end{array}\right\}.
\end{eqnarray}

 In order to compare with the experiment conventionally,
${\cal{M}}^{M_{J_A}M_{J_B}M_{J_C}}(\vec{P})$ can be converted into
the partial amplitude by a recoupling calculation\cite{recp}
\begin{eqnarray}
{\cal{M}}^{LS}(\vec{P})&=&
\sum_{\renewcommand{\arraystretch}{.5}\begin{array}[t]{l}
\scriptstyle M_{J_B},M_{J_C},\\\scriptstyle M_S,M_L
\end{array}}\renewcommand{\arraystretch}{1}\!\!
\langle LM_LSM_S|J_AM_{J_A}\rangle\langle
J_BM_{J_B}J_CM_{J_C}|SM_S\rangle\nonumber\\
&&\times\int
d\Omega\,\mbox{}Y^\ast_{LM_L}{\cal{M}}^{M_{J_A}M_{J_B}M_{J_C}}
(\vec{P}).
\end{eqnarray}
If we consider the relativistic phase space, the decay width
$\Gamma(A\rightarrow BC)$ in terms of the partial wave amplitudes
is
\begin{eqnarray}
\Gamma(A\rightarrow BC)= \frac{\pi
P}{4M^2_A}\sum_{LS}|{\cal{M}}^{LS}|^2. \label{width1}
\end{eqnarray}
Here
$P=|\vec{P}|$=$\frac{\sqrt{[M^2_A-(M_B+M_C)^2][M^2_A-(M_B-M_C)^2]}}{2M_A}$,
$M_A$, $M_B$, and $M_C$ are the masses of the meson $A$, $B$, and
$C$, respectively.

The decay width can be derived analytically if the simple harmonic
oscillator (SHO) approximation for the meson space wave functions
is used. In momentum-space, the SHO wave function is
\begin{eqnarray}
\psi_{nLM_L}(\vec{p})=R^{\mbox{\tiny
SHO}}_{nL}(p)Y_{LM_L}(\Omega_p),
\end{eqnarray}
where the radial wave function is given by
\begin{eqnarray}
R^{\mbox{\tiny SHO}}_{nL}=\frac{(-1)^n(-i)^L}{\beta^{\frac{3}{2}}}
\sqrt{\frac{2n!}{\Gamma(n+L+\frac{3}{2})}}\left(\frac{p}{\beta}\right
)^L e^{-\frac{p^2}{2\beta^2}}L^{L+\frac{1}{2}}_n({\scriptstyle
\frac{p^2}{\beta^2}}).
\end{eqnarray}
Here $\beta$ is the SHO wave function scale parameter, and
$L^{L+\frac{1}{2}}_n({\scriptstyle \frac{p^2}{\beta^2}})$ is  an
associated Laguerre polynomial.

The SHO wave functions cannot be regarded as realistic, however,
they are a {\it {de facto}} standard for many nonrelativistic
quark model calculations. Moreover, the more realistic space wave
functions such as those obtained from Coulomb, plus the linear
potential model do not always result in systematic improvements
due to the inherent uncertainties of the $^3P_0$ decay model
itself\cite{3p0y,3p0x,3p0x2}. The SHO wave function approximation
is commonly employed in the $^3P_0$ decay model in literature. In
the present work, the SHO wave function approximation for the
meson space wave functions is taken.

Under the SHO wave function approximation, the parameters used in
the $^3P_0$ decay model involve the $q\bar{q}$ pair production
strength parameter $\gamma$, the SHO wave function scale parameter
$\beta$, and the masses of the constituent quarks. In the present
work, we take $\gamma=8.77$ and
$\beta_A=\beta_B=\beta_C=\beta=0.4$ GeV, the values recently
obtained by fitting 32 experimentally well-determined decay rates
with the $^3P_0$ decay model\footnote{Our value of $\gamma$ is
higher than that used by Ref.\cite{quarkmass} (0.505) by a factor
of $\sqrt{96\pi}$ due to different field conventions, constant
factor in $T$, etc. The calculated results of the widths are, of
course, unaffected.}, and $m_u=m_d=0.33$ GeV, $m_s=0.55$
GeV\cite{quarkmass}. The meson masses used to determine the phase
space and final state momenta are\footnote{ The assignment the
$K^\ast(1410)$ as the $2\,^3S_1$ kaon is
problematic\cite{3p03,vij}. Quark model\cite{flavorfun} and other
phenomenological approaches\cite{mpla} consistently suggest the
$2\,^3S_1$ kaon has a mass about 1580 MeV, here we take 1580 MeV
as the mass of the $2\,^3S_1$ kaon [$K^\ast(1580)$]. Also, we
assume that the $a_0(1450)$, $K^\ast_0(1430)$, and $f_0(1370)$ are
the ground scalar mesons as Refs.\cite{3p01,3p02,3p03}.}
$M_{\pi}=138$ MeV , $M_K=496$ MeV, $M_{\eta}=548$ MeV,
$M_{\eta^\prime}=958$ MeV, $M_{\rho}=776$ MeV, $M_{K^\ast}=894$
MeV, $M_{\omega}=783$ MeV, $M_{\phi}=1019$ MeV,
$M_{a_2(1320)}=1318$ MeV, $M_{K^\ast_2(1430)}=1429$ MeV,
$M_{f_2(1270)}=1275$ MeV, $M_{f^\prime_2(1525)}=1525$ MeV,
$M_{\pi(1300)}=1240$ MeV, $M_{a_0(1450)}=1474$ MeV,
$M_{K^\ast_0(1430)}=1414$ MeV, $M_{f_0(1370)}=1370$ MeV,
$M_{\rho(1450)}=1459$ MeV, $M_{\omega(1420)}=1420$ MeV,
$M_{K^\ast(1580)}=1580$ MeV, $M_{\rho(1700)}=1720$ MeV,
$M_{K^\ast(1680)}=1717$ MeV, and $M_{K^\ast_3(1780)}=1776$ MeV.

\section{Decays of the $\pi(2070)$}
\indent \vspace*{-1cm}

From (\ref{width1}), the numerical values of the partial decay
widths of the $\pi(2070)$ as the $4\,^1S_0$ isovector state are
listed in Table 1. The initial state mass is set to $2070$ MeV.

\begin{table}[hbt]
\begin{center}
\caption{\small Decays of the $\pi(2070)$ as the $4\,^1S_0$
isovector state in the $^3P_0$ model.}
 \vspace*{0.5cm}
\begin{tabular}{cc|cc}\hline\hline
Mode  & $\Gamma_i$ (MeV)& Mode  & $\Gamma_i$ (MeV)\\\hline
$\rho\omega$ & 3.1&

$\rho\pi$ & 5.9\\

$\pi(1300)\rho$& 52.0&

$\rho(1700)\pi$& 3.6\\

$\rho(1450)\pi$& 112.5&

$ f_2(1270)\pi$ & 48.0\\

$f_0(1370)\pi $ & 0.3&

$ a_2(1320)\eta$ & 8.8\\

$ a_0(1450)\eta$ & 5.1&

$K K^\ast$ & 8.5\\

$ K^\ast K^\ast$ & 14.9&

$ K^\ast_0(1430)K$& 10.3\\

$ K^\ast_2(1430)K$ & 4.6& \\\hline

\multicolumn{4}{c}{$\Gamma_{\mbox{thy}}=277.6$ MeV,
$\Gamma_{\mbox{expt}}=310^{+100}_{-50}$ MeV}\\
         \hline\hline
\end{tabular}
\end{center}
\end{table}

\begin{figure}[hbt]
\begin{center}
\epsfig{file=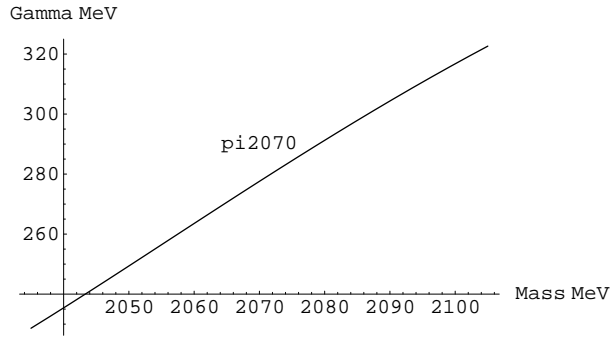,width=8.0cm, clip=}
\vspace*{0.5cm}\vspace*{-1cm}
 \caption{\small The predicted total width of the $\pi(2070)$ as the $4\,^1S_0$ isovector versus the inital state mass.}
\end{center}
\end{figure}

Table 1 indicates that the total width of the $\pi(2070)$ as the
$4\,^1S_0$ isovector state predicted by the $^3P_0$ decay model is
about $278$ MeV, consistent with the observation
$(310^{+100}_{-50})$ MeV within errors, and the dominant decay
modes are expected to be $\pi(1300)\rho$, $\rho(1450)\pi$ and $
f_2(1270)\pi$. Also, in order to check the dependence of the
theoretical result on the initial state mass, the predicted total
width of the $\pi(2070)$ is shown in Fig. 2 as the function of the
initial state mass. Fig. 2 shows that when the initial state mass
varies from 2035 to 2105 MeV, the total width of the $4\,^1S_0$
isovector state varies from about 230 to 320 MeV, generally in
accord with the width range of the $\pi(2070)$. Both the mass and
width of the $\pi(2070)$ are consistent with the predicted
$4\,^1S_0$ isovector state, which therefore suggests that the
assignment of the $\pi(2070)$ as the $4\,^1S_0$ isovector state
seams plausible.

\section{Decays of the $\eta(2010)$, $\eta(2100)$, $\eta(2190)$, and $\eta(2225)$}
\indent \vspace*{-1cm}

 In the presence of the $\eta(2225)$ being one isoscalar member of the $4\,^1S_0$
meson nonet, the $\eta(2010)$, $\eta(2100)$, and $\eta(2190)$
would complete another isoscalar member. It is well known that in
a meson nonet, the two physical isoscalar states can mix. The
mixing of the two isoscalar states can be parameterized as
\begin{eqnarray}
&&\eta(x)=\cos\phi~ n\bar{n}-\sin\phi~ s\bar{s},\\
&&\eta(2225)=\sin\phi~ n\bar{n}+\cos\phi~ s\bar{s},
\end{eqnarray}
where $n\bar{n}=(u\bar{u}+d\bar{d})/\sqrt{2}$ and $s\bar{s}$ are
the pure $4\,^1S_0$ nonstrange and strange states, respectively,
and $\eta(x)$ denotes the $\eta(2010)$, $\eta(2100)$, or
$\eta(2190)$.

According to (\ref{width1}), the partial widths of $\eta(x)$ and
$\eta(2225)$ become with mixing
\begin{eqnarray}
&&\Gamma(\eta(x)\rightarrow
BC)=\frac{\pi~P}{4M^2_{\eta(x)}}\sum_{LS}|\cos\phi
{\cal{M}}^{LS}_{n\bar{n}\rightarrow BC}-\sin\phi
{\cal{M}}^{LS}_{s\bar{s}\rightarrow BC}|^2,
\label{w1}\\
 &&\Gamma(\eta(2225)\rightarrow
BC)=\frac{\pi~P}{4M^2_{\eta(2225)}}\sum_{LS}|\sin\phi
{\cal{M}}^{LS}_{n\bar{n}\rightarrow BC}+\cos\phi
{\cal{M}}^{LS}_{s\bar{s}\rightarrow BC}|^2. \label{w2}
\end{eqnarray}
Based on (\ref{w1}) and (\ref{w2}), the predicted total widths of
the $\eta(2010)$, $\eta(2100)$, $\eta(2190)$, and $\eta(2225)$ are
shown in Fig. 3 as functions of the initial state mass and the
mixing angle $\phi$.  From Fig. 3, one can see that with the
variations of the initial state mass and $\phi$, only the measured
widths of the $\eta(2100)$ and $\eta(2225)$ are possible to be
reasonably reproduced in the $^3P_0$ model. We therefore suggest
that the assignment of the $\eta(2010)$ and $\eta(2190))$ as the
$4\,^1S_0$ isoscalar states seems unfavorable. We shall focus on
the possibility of the $\eta(2100)$ being the partner of the
$\eta(2225)$. Taking $M_{\eta(2100)}=2103$ MeV and
$M_{\eta(2225)}=2240$ MeV, we list the numerical values of the
partial decay widths of the $\eta(2100)$ and $\eta(2225)$ in Table
2. The variation of the theoretical total widths of the
$\eta(2100)$ and $\eta(2225)$ with the mixing angle $\phi$ is
shown in Fig. 4.

\begin{figure}[hbt]
\begin{center}
\epsfig{file=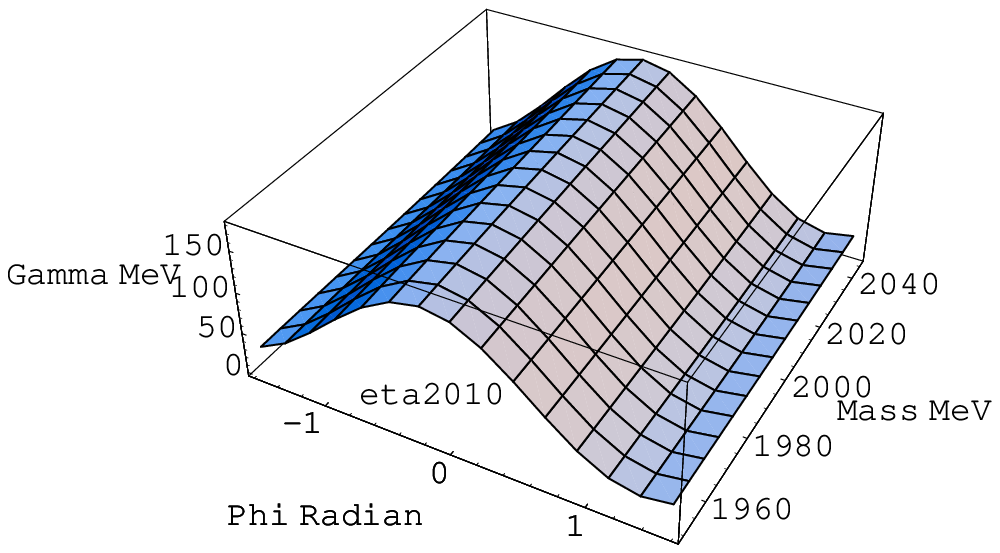,width=6.0cm, clip=}
\epsfig{file=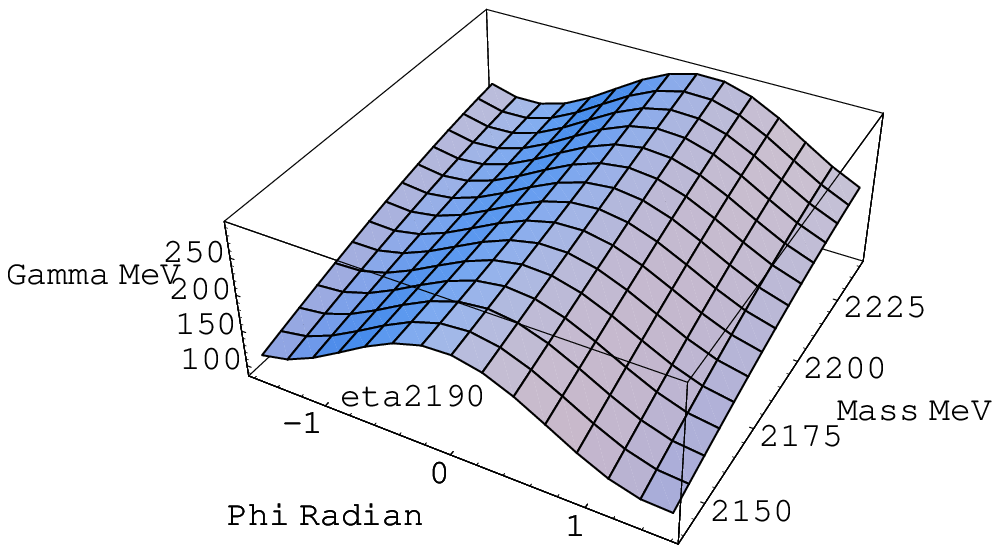,width=6.0cm, clip=}
\epsfig{file=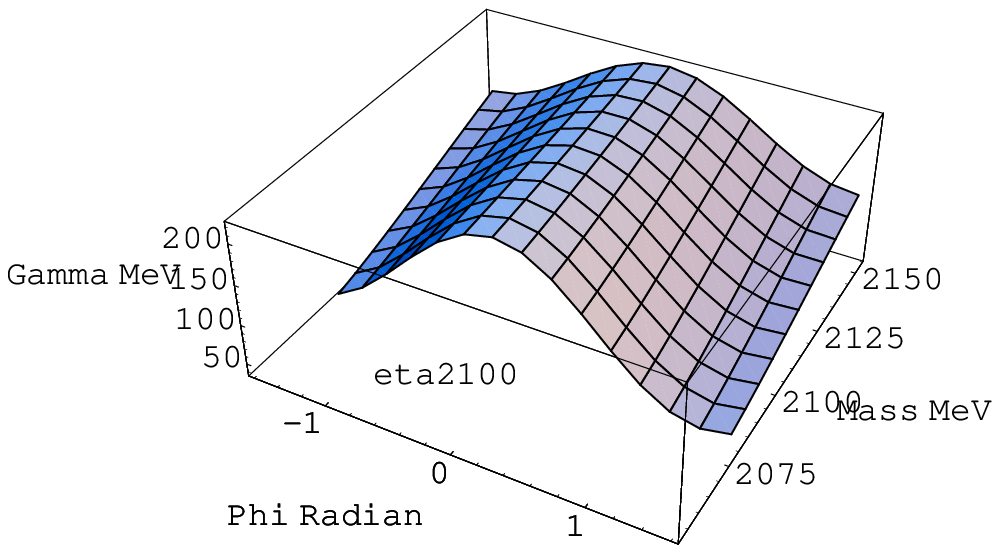,width=6.0cm, clip=}
\epsfig{file=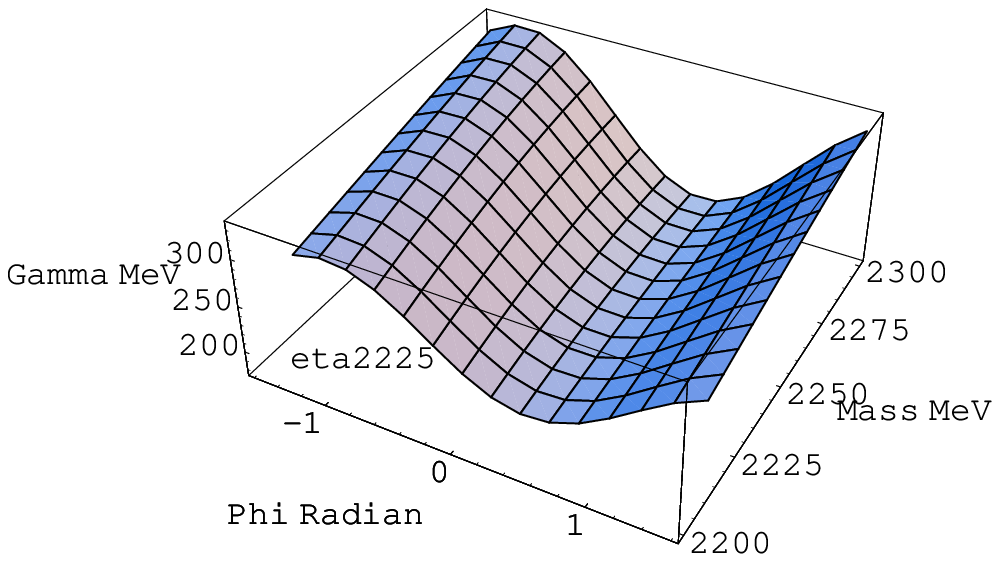,width=6.0cm, clip=}
\vspace*{0.5cm}\vspace*{-1cm}
 \caption{\small The predicted total widths of the $\eta(2010)$, $\eta(2190)$, $\eta(2100)$, and $\eta(2225)$ as the $4\,^1S_0$ isoscalar states versus the inital state mass and the mixing angle $\phi$.}
\end{center}
\end{figure}

\begin{table}[hbt]
\begin{center}
\caption{\small Decays of the $\eta(2100)$ and $\eta(2225)$ as the
$4\,^1S_0$ isoscalar states in the $^3P_0$ model.
$c\equiv\cos\phi$, $s\equiv\sin\phi$.} \vspace*{0.5cm}
\begin{tabular}{c|c|c}\hline\hline
   \multicolumn{1}{c}{} &\multicolumn{1}{c}{$\eta(2100)$}& \multicolumn{1}{c}{$\eta(2225)$}
         \\  \hline
   Mode  & $\Gamma_i$(MeV) &    $\Gamma_i$(MeV)
   \\\hline
$\rho\rho$ &
  $2.1c^2$&$1.4s^2$\\
$\omega\omega$&
  $0.9c^2$&$0.3s^2$\\
$\phi\phi$& $9.7s^2$
  &$20.1c^2$\\
$a_2(1320)\pi$&
  $143.7c^2$&$158.8s^2$\\
$a_0(1450)\pi$&
 $0.1c^2$&$4.5s^2$\\
$KK^\ast$&
  $7.4c^2-15.0cs+7.6s^2$&$14.4c^2+12.7cs+2.8s^2$\\
$K^\ast K^\ast$& $15.4c^2-13.4cs+2.9s^2$
  &$0.8c^2-6.6cs+13.6s^2$\\
$KK^\ast_0(1430)$&
  $8.8c^2-7.9cs+1.8s^2$&$2.5c^2-4.7cs+2.2s^2$\\
$K K^\ast_2(1430)$& $7.7c^2-31.1cs+31.3s^2$
  &$69.4c^2+91.9cs+30.4s^2$\\
$KK^\ast(1580)$&
  $5.4c^2+17.2cs+13.6s^2$&$90.7c^2-158.1cs+68.9s^2$\\
$K K^\ast(1680)$&
  &$1.6c^2-1.4cs+0.3s^2$\\
  \hline
&\multicolumn{1}{c}{$\Gamma_{\mbox{thy}}=191.5c^2-50.2cs+66.9s^2$}&
\multicolumn{1}{|c}{$\Gamma_{\mbox{thy}}=199.4c^2-66.2cs+283.2s^2$}\\
&\multicolumn{1}{c}{$\Gamma_{\mbox{expt}}=187 \pm 75$}&
\multicolumn{1}{|c}{$\Gamma_{\mbox{expt}}=190\pm 30^{+40}_{-60}$}
         \\  \hline
 \hline
\end{tabular}
\end{center}
\end{table}

\begin{figure}[hbt]
\begin{center}
\epsfig{file=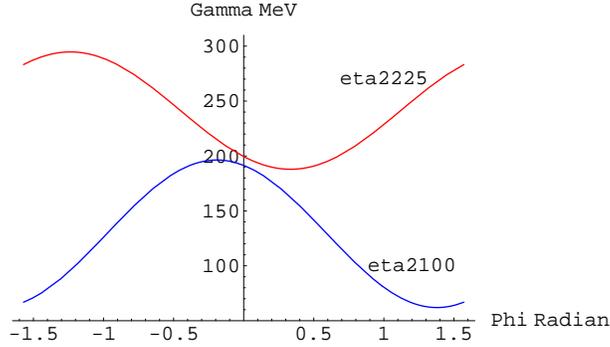,width=8.0cm, clip=}
\vspace*{0.5cm}\vspace*{-1cm}
 \caption{\small The predicted total widths of the $\eta(2100)$ and $\eta(2225)$ as the $4\,^1S_0$ isoscalar states versus the mixing angle $\phi$.}
\end{center}
\end{figure}

From Fig. 4, we find that if the $\eta(2100)-\eta(2225)$ mixing
angle $\phi$ lying in the range from about $-0.6$ to $+0.7$
radians, both the measured widths of the $\eta(2100)$ and
$\eta(2225)$ can be reasonably reproduced. In order to check
whether the possibility of $-0.6 \leq \phi \leq +0.7$ radians
exists or not, below we shall estimate the
$\eta(2100)$-$\eta(2225)$ mixing angle phenomenologically.

 In
the $n\bar{n}$ and $s\bar{s}$ basis, the mass-squared matrix
describing the $\eta(2100)$ and $\eta(2225)$ mixing can be written
as\cite{jpglidm,epja-1}
\begin{eqnarray}
M^2=\left(\begin{array}{cc}
M^2_{n\bar{n}}+2A_m&\sqrt{2}A_mX\\
\sqrt{2}A_mX&M_{s\bar{s}}^2+A_mX^2
\end{array}\right),
\label{matrix}
\end{eqnarray}
where $M_{n\bar{n}}$ and $M_{s\bar{s}}$ are the masses of the
states $n\bar{n}$ and $s\bar{s}$, respectively, $A_m$ denotes the
total annihilation strength of the $q\bar{q}$ pair for the light
flavors $u$ and $d$, $X$ describes the $SU(3)$-breaking ratio of
the nonstrange and strange quark masses via the constituent quark
mass ratio $m_u/m_s$. The masses of the two physical states
$\eta(2100)$ and $\eta(2225)$ can be related to the matrix $M^2$
by the unitary matrix $U=\left(\begin{array}{cc}
\cos\phi&-\sin\phi\\
\sin\phi&\cos\phi\end{array}\right)$
\begin{eqnarray}
U M^2 U^\dagger=\left(\begin{array}{cc}
M^2_{\eta(2100)}&0\\
0&M^2_{\eta(2225)}\end{array}\right). \label{diag}
\end{eqnarray}
$n\bar{n}$ is the orthogonal partner of the $\pi(4^1S_0)$, the
isovector state of $4\,^1S_0$ meson nonet, and one can expect that
$n\bar{n}$ degenerates with $\pi(4\,^1S_0)$ in effective quark
masses, here we take $M_{n\bar{n}}=M_{\pi(4^1S_0)}=M_{\pi(2070)}$.
With the help of the Gell-Mann-Okubo mass formula
$M^2_{s\bar{s}}=2M^2_{ K(4^1S_0)}-M^2_{n\bar{n}}$\cite{okubo}, the
following relations can be derived from (\ref{diag})
\begin{eqnarray}
8X^2(M^2_{K(4^1S_0)}-M^2_{\pi(2070)})^2&=&[4M^2_{K(4^1S_0)}-(2-X^2)M^2_{\pi(2070)}-(2+X^2)M^2_{\eta(2100)}]\nonumber\\
&&\times[(2-X^2)M^2_{\pi(2070)}+(2+X^2)M^2_{\eta(2225)}-4M^2_{K(4^1S_0)}],
\label{msch}
\end{eqnarray}
\begin{eqnarray}
A_m=\frac{(M^2_{\eta(2225)}-2M^2_{K(4^1S_0)}+M^2_{\pi(2070)})(M^2_{\eta(2100)}-2M^2_{K(4^1S_0)}+M^2_{\pi(2070)})}{2(M^2_{\pi(2070)}-M^2_{K(4^1S_0)})X^2}.
\label{am}
\end{eqnarray}
If the $SU(3)$-breaking effect is not considered, i.e., $X=1$,
relation (\ref{msch}) can be reduced to Schwinger's original nonet
mass formula\cite{schwinger}. Taking $X=m_u/m_s=0.33/0.55=0.6$,
from (\ref{msch}) and (\ref{am}) we have
\begin{eqnarray}
M_{K(4\,^1S_0)}=2.153~ \mbox{GeV},~ A_m=0.07~ {\mbox{GeV}}^2.
\end{eqnarray}

Based on the values of the above parameters involved in
(\ref{matrix}), the unitary matrix $U$ can be given by
\begin{eqnarray}
U=\left(\begin{array}{cc}
\cos\phi&-\sin\phi\\
\sin\phi&\cos\phi\end{array}\right)=\left(\begin{array}{cc}
 +0.995&-0.104\\
+0.104&+0.995
\end{array}\right),
\label{mixangle}
\end{eqnarray}
which gives $\phi= +0.1$ radians, just lying in the range from
about $-0.6$ to $+0.7$ radians.  From Table 2, this estimated
mixing angle leads to $\Gamma_{\mbox{thy}}(\eta(2100))=185.2$ MeV
and $\Gamma_{\mbox{thy}}(\eta(2225))=193.7$ MeV, both in good
agreement with the experimental results. The $\eta(2100)$ and
$\eta(2225)$, together with the $\pi(2070)$ therefore appear to be
the convincing $4\,^1S_0$ states.

\section{ Decays of the $4\,^1S_0$ kaon}
\indent \vspace*{-1cm}

The above two sections show that in the presence of the
$\pi(2070)$, $\eta(2100)$, and $\eta(2225)$ belonging to the
$4\,^1S_0$ meson nonet, the total widths of these three states can
be naturally accounted for in the $^3P_0$ decay model, and the
$4\,^1S_0$ kaon is expected to have a mass of about 2153 MeV by
the mass formula (\ref{msch}). Below the $K(2150)$ denotes the
$4\,^1S_0$ kaon. We note that the $K$, $K(1460)$\footnote{The
$K(1460)$ mass is taken 1400 MeV reported by\cite{k1460}.},
$K(1830)$, and $K(2150)$ approximately populate a common
trajectory as shown in Fig. 5. The quasi-linear trajectories at
the (n, Mass-squared)-plots turned out to be able to describe the
light mesons with a good accuracy\cite{reggeph}. Fig. 5 therefore
indicates that the $K(1460)$, $K(1830)$, and $K(2150)$ could be
the good candidates for the $2\,^1S_0$, $3\,^1S_0$, and $4\,^1S_0$
kaons, respectively.

\begin{figure}[hbt]
\begin{center}
\epsfig{file=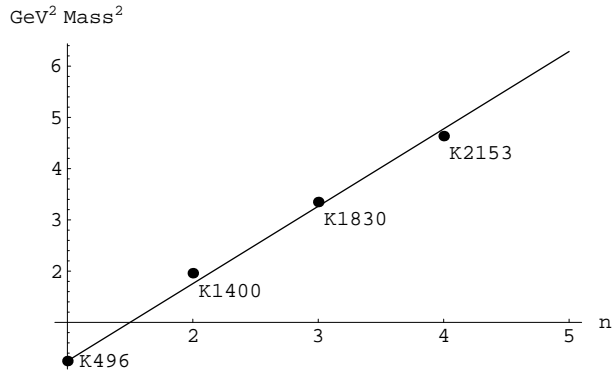,width=8.0cm, clip=}
\vspace*{0.5cm}\vspace*{-1cm}
 \caption{\small The (n, Mass-squared)-trajectory for the $K$.}
\end{center}
\end{figure}

The $K(2150)$ is not related any current experimental candidate.
The predicted decay widths of the $K(2150)$ are listed in Table 3.
The initial state mass is set to 2153 MeV. The total width of the
$K(2150)$ is predicted to be about 197 MeV, and the dominant decay
modes of the $K(2150)$ are expected to be $ K^\ast_2(1430)\pi$, $
K^\ast(1580)\pi$, $\rho(1459)K$ and $a_2(1320)K$. These results
could be of use in searching the candidate for the $4\,^1S_0$ kaon
experimentally.

\begin{table}[hbt]
\begin{center}
\caption{\small Decays of the $K(2150)$ as the $4\,^1S_0$
isodoublet in the $^3P_0$ model.} \vspace*{0.5cm}
\begin{tabular}{cc|cc}\hline\hline
Mode  & $\Gamma_i$ (MeV)&Mode  & $\Gamma_i$ (MeV)\\\hline

$\rho K$ & 3.9&

$\omega K$ & 1.3\\

$\phi K$ & 1.3&

$\rho K^\ast$ & 0.08\\

$\omega K^\ast$& 0.04&

$\phi K^\ast$ & 12.0\\

$\pi K^\ast$ & 4.5&

$\eta K^\ast$& 1.5\\

$\eta^\prime K^\ast$& 0.09&

$ K^\ast_0(1430)\pi$& 1.6\\

$ K^\ast_2(1430)\pi$& 29.7&

$ K^\ast(1580)\pi$& 34.5\\

$ K^\ast(1680)\pi$& 2.5&

$ K^\ast_3(1780)\pi$& 1.0\\

$\pi(1300) K^\ast$& 6.3&

$\rho(1450) K$& 42.8\\

$\omega(1420) K$& 14.6&

$a_2(1320) K$& 26.7\\

$f_2(1270)K$ & 9.9&

$ f^\prime_2(1525) K$& 2.9\\

\hline

\multicolumn{4}{c}{$\Gamma_{\mbox{thy}}=197.2$ MeV}\\
         \hline\hline
\end{tabular}
\end{center}
\end{table}

\section{Summary and conclusion}
\indent \vspace*{-1cm}

With the assignment of the $\eta(2225)$ recently observed by the
BES Collaboration as the $s\bar{s}$ member of the $4\,^1S_0$ meson
nonet, the possibility of the $\pi(2070)$, $\eta(2010)$,
$\eta(2100)$, and $\eta(2190)$ being the $4\,^1S_0$ $q\bar{q}$
states is discussed. With respect to the $\pi(2070)$, its
assignment to the $4\,^1S_0$ isovector state is not only favored
by its mass, but also by its width. The assignment of the
$\eta(2010)$ and $\eta(2190)$ as the $4\,^1S_0$ isoscalar states
is not favored by their widths. Both the widths of the
$\eta(2100)$ and $\eta(2225)$ can be reasonably reproduced with
the mixing angle lying in the range  from about $-0.6$ to $+0.7$
radians. The assignment of the $\pi(2070)$, $\eta(2100)$, and
$\eta(2225)$ as the members of the $4\,^1S_0$ meson nonet not only
leads to that the $\eta(2100)$-$\eta(2225)$ mixing angle is about
$+0.1$ radians which naturally accounts for the widths of the
$\eta(2100)$ and $\eta(2225)$, but also gives that the $4\,^1S_0$
kaon has a mass of about 2153 MeV. The $K$, $K(1460)$, $K(1830)$,
and $K(2150)$ approximately populate a common (n,
Mass-squared)-trajectory.  We tend to conclude that the observed
pseudoscalar states $\pi(2070)$, $\eta(2100)$, $\eta(2225)$,
together with the unobserved $K(2150)$ appear to be the good
candidates for the members of the $4\,^1S_0$ meson nonet. The
$K(2150)$ width is predicted to be about 197 MeV, and the dominant
decay modes of the $K(2150)$ are expected to be $
K^\ast_2(1430)\pi$, $ K^\ast(1580)\pi$, $\rho(1459)K$ and
$a_2(1320)K$. These results could be of use in searching the
candidate for the $4\,^1S_0$ kaon experimentally.

 \section*{Acknowledgments}
 This work
is supported in part by HANCET under Contract No. 2006HANCET-02,
and the Program for Youthful Teachers in University of Henan
Province.

\end{document}